\documentclass[fleqn,usenatbib, a4paper, margin=2.0cm]{mnras}
\usepackage{supertabular,booktabs}

\usepackage{enumitem,xcolor,multicol,amssymb}
\usepackage{colortbl,array,natbib,graphicx,amsmath}
\usepackage{threeparttable}
\usepackage{verbatimbox}
\usepackage{longtable}
\usepackage{newtxtext,newtxmath}
\usepackage{siunitx}
\usepackage{pdflscape}

\usepackage[T1]{fontenc}

\newcolumntype{C}[1]{>{\centering\let\newline\\\arraybackslash\hspace{0pt}}m{#1}}
\DeclareRobustCommand{\VAN}[3]{#2}
\let\VANthebibliography\thebibliography
\def\thebibliography{\DeclareRobustCommand{\VAN}[3]{##3}\VANthebibliography}
\usepackage{graphicx}	
\usepackage{amsmath}	
\usepackage{amssymb}	

\title[GC ULXs in the Furthest Early-Type Galaxies]{Globular Cluster Ultraluminous X-ray Sources in the Furthest Early-Type Galaxies}

\author[E. Thygesen, et al.]{
Erica Thygesen,$^{1}$\thanks{E-mail: thygesen@msu.edu}, Yifan Sun$^{2,3}$, Jeff Huang$^{2,3}$ ,  Kristen C. Dage$^{2,3}$, Stephen E. Zepf$^{1}$,
\newauthor
Arunav Kundu$^{4}$, Daryl Haggard$^{2,3}$,  Thomas J. Maccarone$^{5}$ 
\\
\\
$^{1}$Department  of  Physics  and  Astronomy,  Michigan  State  University,  East Lansing, MI 48824, USA\\
$^{2}$ Department of Physics, McGill University, 3600 University Street, Montr\'eal, QC H3A 2T8, Canada\\
$^{3}$ McGill Space Institute, McGill University, 3550 University Street, Montr\'eal, QC H3A 2A7, Canada\\
$^{4}$ Eureka Scientific, Inc., 2452 Delmer Street, Suite 100 Oakland, CA 94602, USA\\
$^{5}$ Department of Physics \& Astronomy, Box 41051, Science Building, Texas Tech University, Lubbock, TX 79409-1051, USA \\}
\date{Accepted XXX. Received YYY; in original form ZZZ}

\pubyear{2021}

\begin{document}
\maketitle

\begin{abstract}
Ultraluminous X-ray Sources (ULXs) in globular clusters are low mass X-ray binaries that achieve high X-ray luminosities through a currently uncertain accretion mechanism. Using archival \textit{Chandra} and \textit{Hubble Space Telescope} (\textit{HST}) observations, we perform a volume-limited search ($\lesssim$ 70 Mpc) of 21 of the most massive ($>10^{11.5} M_\odot$) early-type galaxies to identify ULXs hosted by globular cluster (GC) candidates. We find a total of 34 ULX candidates above the expected background within 5 times the effective radius of each galaxy, with 10 of these ($\sim29.4\%$) potentially hosted by a GC. A comparison of the spatial and luminosity distributions of these new candidate GC ULXs with previously identified GC ULXs shows that they are similar: both samples peak at $L_X \sim$ a few $\times 10^{39}$ erg/s and are typically located within a few effective radii of their host galaxies.
\end{abstract}


\section{Introduction}
Ultraluminous X-ray sources (ULXs) are bright, off-nuclear X-ray point sources that exceed the Eddington limit for a 10 $M_\odot$ black hole (BH): $L_X \gtrsim10^{39}$ erg s$^{-1}$ \citep{fabbiano89}. While many ULXs were initially thought to be evidence for elusive intermediate mass BHs (IMBHs), it is quite likely that the best explanation for most ULXs is super-Eddington accretion onto a stellar mass compact object, either a neutron star (NS; \citealt{2014Natur.514..202B}) or a BH \citep{2007Natur.445..183M}, although some are still tenable IMBH candidates (e.g., HLX-1; \citealt{Farrell09}). Many ULXs are now being observed to exhibit pulsations \citep{2014Natur.514..202B}, confirming the existence of NS accretors in these systems. Thus, ULXs are a heterogeneous class of objects.

Most ULXs have been identified within star-forming regions of spiral galaxies, but a very different population of ULXs is emerging in globular clusters (GCs; e.g., \citealt{2007Natur.445..183M,2021MNRAS.504.1545D}). GCs are dense groups of primarily old stars -- a stark contrast to the more typical host environments of ULXs. The crowded nature of globular clusters means that X-ray binaries in these environments are more likely to form via dynamical channels, rather than in situ (as is typical with field ULXs). X-ray binaries in GCs will also likely have low-mass companions due to the age of these stellar populations. Thus, the nature of ULXs in globular clusters is very different than those found in spiral galaxies, both on account of the binary make up as well as the formation history. 

Given that GC ULXs may be indicative of BH accretors (e.g., \citealt{2007Natur.445..183M,2018ApJ...862..108D}), probing the local volume for GC ULXs informs on the observable number of BHs in GCs. Early theoretical work \citep{1969ApJ...158L.139S,1993Natur.364..423S,1993Natur.364..421K} suggested that BHs would frequently be ejected from their host GCs due to gravitational interaction, but more recent theoretical and observational studies have shown that a significant fraction of BHs are actually retained \citep{2009ApJ...690.1370M, 2015MNRAS.453.3918M, 2018MNRAS.475L..15G, 2019A&A...632A...3G}. Current theoretical simulations \citep{2015ApJ...800....9M,2018ApJ...852...29K} are also suggesting that BH-BH binaries are formed in GCs, making the question of BHs in GCs pertinent in today's multi-messenger universe. Beyond this, the recent discovery of a fast radio burst localised to a globular cluster in M81 by \cite{2022Natur.602..585K} highlights the importance of studying peculiar sources in globular clusters.

The nature of GC ULXs and their host clusters is not yet well-understood. The known population of GC ULXs is currently in its infant stages, with only 20 candidates to date \citep[and references therein]{2021MNRAS.504.1545D} and none in Galactic GCs. Since all known GC ULXs are extragalactic, there is an inherent challenge of telescope sensitivity detecting the ULXs and their parent clusters, thus limiting us to nearby searches or, for more distant galaxies, only the brightest sources. The most massive early-type galaxies can host up to thousands of globular clusters, making them ideal targets for large-scale surveys directed at increasing the known population of GC ULXs. While studies of GC ULXs in a handful of nearby elliptical galaxies have been conducted, a statistical sample of GC ULXs in early-type galaxies is still being built. 

In this study, we seek to increase the number of candidate GC ULXs by searching massive, early-type galaxies. We capitalise on a previous survey of galaxies of this type (MASSIVE Survey, \citealt{2014ApJ...795..158M}) that was aimed at studying their diffuse X-ray properties using the \textit{Chandra} X-Ray Observatory. Using distant galaxies for this purpose has both pros and cons -- the downside being the low signal-to-noise for many of the images, but the upside is that we are inherently limiting ourselves to the brightest sources, so our images are not affected by contamination from low-luminosity sources. Here we couple the X-ray data with images from the Hubble Source Catalogue (HSC) to identify ULXs potentially associated with GCs, which significantly reduces the chances that the bright X-ray luminosities are due to background AGN \citep{2007Natur.445..183M}. 

In this paper we probe the furthest distances to detect GC ULXs in early-type galaxies. The data and analysis methods are laid out in Section \ref{sec:data} and the population results are discussed in Section \ref{sec:results}. We provide a brief summary for each galaxy in Section \ref{sec:gal_summaries}. Finally, we discuss the implications of this work in Section \ref{section:discusson}. 

\section{Data and Analysis}
\label{sec:data}
In this paper, we examine X-ray and optical data from \textit{Chandra} and \textit{HST} respectively to identify  candidate globular cluster ULXs in 21 early-type galaxies. The data and analysis methods are outlined below.

\subsection{Sample Selection}\label{sec:sample_selection}

To capitalise on existing archival data, we select our sample of galaxies from the MASSIVE Survey \citep{2014ApJ...795..158M}, which is a volume- and magnitude-limited survey of some of the most massive ($M_* \geq 10^{11.5} M_\odot$) early-type galaxies. We limit our search to $\leq 70$ Mpc so that data quality is not sacrificed. The three nearest galaxies ($\leq 16.7$ Mpc) from the MASSIVE survey have been studied previously for their GC ULX population (NGC 4649, NGC 4472 and NGC 4486/M87; \citealt{2019MNRAS.485.1694D}; \citealt{2020MNRAS.497..596D}), so we only include the galaxies that are further than these (34.2 Mpc $\leq$ D $\leq$ 70 Mpc). Our sample consists of the remaining 21 galaxies from this selection (Table \ref{tab:gals}). Three galaxies did not have sufficient archival \textit{HST} data (NGC 128, NGC 499 and NGC 1066), and three only had central X-ray emission which would be inconsistent with a ULX classification (NGC 1453, NGC 2258 and NGC 2672).

\begin{table}
\caption{Final galaxy sample derived from the MASSIVE Survey, selected based on having available \textit{Chandra} observations.}
\begin{threeparttable}
\begin{tabular}{lccccc}
\hline\hline
Galaxy & RA & Dec & Dist &  Log$(M_*)$ & $R_\mathrm{eff}$   \\
   & & & (Mpc) &  ($M_\odot$) & (arcsec) \\
\hline
NGC-128 & 00:29:15.07 & +02:51:50.76 & 59.3 & 11.61 & 10.5 \\
NGC-499 & 01:23:11.47 & +33:27:36.36 & 69.8 & 11.87 & 11.6 \\
NGC-507 & 01:23:39.94 & +33:15:21.96 & 69.8 & 11.87 & 23.0 \\
NGC-708 & 01:52:46.49 & +36:09:06.48 & 69.0 & 11.75 & 23.7 \\
NGC-890 & 02:22:01.01 & +33:15:57.96 & 55.6 & 11.68 & 16.7 \\
NGC-1060 & 02:43:15.05 & +32:25:30.00 & 67.4 & 11.90 & 16.8 \\
NGC-1066 & 02:43:49.90 & +32:28:29.64 & 67.4 & 11.60 & 17.5 \\
NGC-1453 & 03:46:27.26 & -03:58:07.68 & 56.4 & 11.75 & 16.0 \\
NGC-1573 & 04:35:03.98 & +73:15:44.64 & 65.0 & 11.70 & 13.9 \\
NGC-1600 & 04:31:39.86 & -05:05:09.96 & 63.8 & 11.90 & 20.8 \\
NGC-1684 & 04:52:31.15 & -03:06:21.96 & 63.5 & 11.61 & 15.8 \\
NGC-1700 & 04:56:56.33 & -04:51:56.88 & 54.4 & 11.72 & 13.4 \\
NGC-2258 & 06:47:46.20 & +74:28:54.48 & 59.0 & 11.75 & 18.6 \\
NGC-2672 & 08:49:21.89 & +19:04:30.00 & 61.5 & 11.72 & 16.9 \\
NGC-5322 & 13:49:15.19 & +60:11:25.44 & 34.2 & 11.68 & 26.6 \\
NGC-5353 & 13:53:26.71 & +40:16:59.16 & 41.1 & 11.66 & 14.2 \\
NGC-5557 & 14:18:25.70 & +36:29:36.96 & 51.0 & 11.66 & 16.2 \\
NGC-6482 & 17:51:48.82 & +23:04:18.84 & 61.4 & 11.72 & 10.1 \\
NGC-7052 & 21:18:33.05 & +26:26:48.84 & 69.3 & 11.75 & 14.7 \\
NGC-7619 & 23:20:14.52 & +08:12:22.68 & 54.0 & 11.75 & 14.8 \\
NGC-7626 & 23:20:42.53 & +08:13:01.20 & 54.0 & 11.75 & 20.1 \\
\hline
\end{tabular}
\end{threeparttable} 
\label{tab:gals} 
\end{table}

\subsection{\textit{Chandra} X-Ray Observations}\label{sec:chandra}
\label{subsec:chandra}
The archival \textit{Chandra} data are reduced using the \textit{Chandra} Interactive Analysis of Observations ({\sc ciao}) software version 4.10 \citep{2006SPIE.6270E..1VF}. The \textit{Chandra} observations were obtained with both ACIS-S and ACIS-I instruments. Information on the observations is tabulated in the Appendix (Table \ref{tab:obs}). To identify point sources, we run {\tt wavdetect} on the full-band images with the sigthresh value set to $\sim10^{-6}$, corresponding to approximately one false positive per chip. The enclosed count fraction is set to 0.9. The wavelet scales are set to 1.0, 2.0, 4.0, 8.0, 16.0, and the point spread function (PSF) map energy is set to 2.3 keV. Due to the generally high amount of central X-ray gas in these galaxies, {\tt wavdetect} can erroneously identify sources within the inner regions of the galaxies, which we remove from our source list.

We attempt to match the \textit{Chandra} astrometry to \textit{Gaia} sources, however, there are not enough high-quality matches in our images to perform this correction. We therefore assume the absolute astrometry of \textit{Chandra} to be 0.8". To find the 95\% positional uncertainty of the
X-ray sources we use Equation 5 from \citealt{2005ApJ...635..907H}, which determines the error circle for sources identified by {\tt wavdetect} as a function of source counts and distance between the \textit{Chandra} aimpoint and the location of the source on the detector. We add this in quadrature with the 0.8" \textit{Chandra} uncertainty to create the final positional uncertainties, although adding this value is negligible. Any optical source within this area of uncertainty is considered as a match to the X-ray source.

We run {\tt srcflux} on the soft (0.5-2 keV), hard (2-8 keV) and full (0.5-8 keV) energy bands to calculate the fluxes of all {\tt wavdetect} sources in our images. We generate background regions for each source using {\tt roi}. We use a 90\% confidence interval for the flux calculations, and assume a fixed power law model with a photon index of $\Gamma=1.7$ to calculate the unabsorbed flux. Galactic absorption for each source is taken into account using {\tt colden}. 

As we are only interested in the brightest sources, we limit our sample to sources with unabsorbed $L_{0.5-8 \mathrm{ keV}} \geq 7\times10^{38}$ erg s$^{-1}$ in at least one ObsID. We report the luminosity of the candidate ULXs even if they are detected below this threshold in other ObsIDs (see Table \ref{tabA2:sources}). As a first test for association with the galaxy, we only select bright sources within five times the effective radius ($R_\mathrm{eff}$; see Equation 3 from \citealt{2014ApJ...795..158M}) of each galaxy for further investigation. 

To account for the probability of the X-ray sources being background objects, we use Equation 2 of \cite{2008MNRAS.388.1205G} to calculate the number of X-ray sources expected to be observed above a given flux within a given area, accounting for various observational effects. Our fluxes are calculated for 0.5-8 keV, but we use the background source estimation for 0.5-10 keV as the flux at the 8-10 keV range is negligible. Most of the X-ray positional uncertainties are $\leq 1"$, which has an expected background source count of $\sim 3.4\times10^{-4}$ at this flux interval. As a further test for reliability, we find a significant excess in the number of observed sources compared to the number of expected background sources within various multiples of the effective radius of each galaxy (Table \ref{tab:bkg}).

\subsection{HST Optical Observations}

We obtain optical and infrared measurements from the Hubble Source Catalog (HSC) version 3.1. We query all sources within each galaxy with \textit{HST} filters F555W (WFPC2), F814W (WFPC2 or ACS) and F110W (WFC3). Fifteen of the galaxies in this sample have all been observed in the F110W filter (see Table \ref{tab:gals}), while six galaxies have no \textit{HST} observations. The astrometry of HSC is already calibrated using catalogues such as \textit{Gaia} DR1. We cross-match the ULX candidates to the F110W point source catalogues, using the X-ray positional uncertainties as the matching radius.

\subsection{Classification of Optical Counterparts}

\begin{figure}
    \centering
    \includegraphics[width=0.5\textwidth]{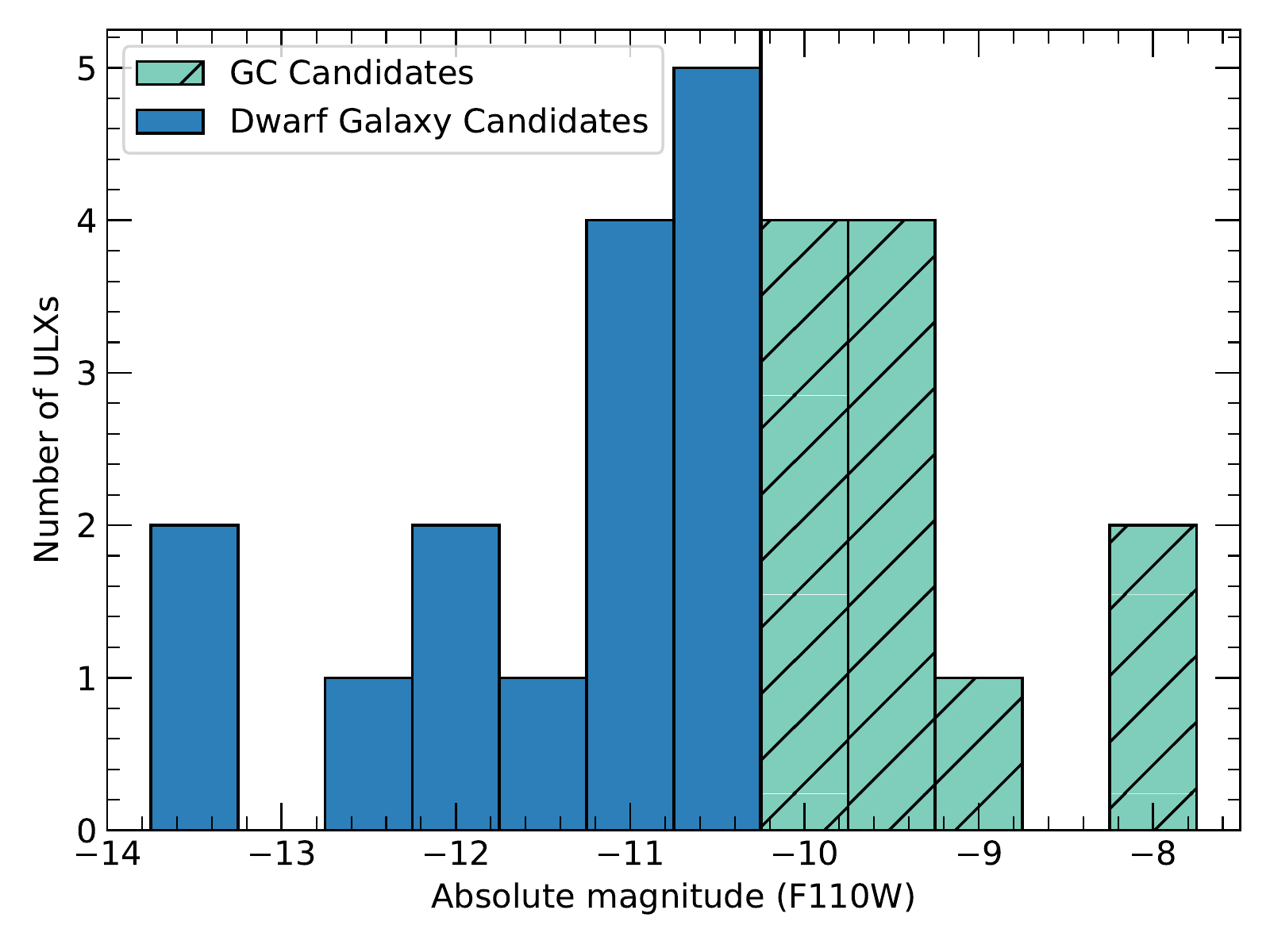}
    \caption{Histogram of cross-matched HSC sources binned by their F110W absolute magnitudes. Sources in blue have magnitudes brighter than -10.25, and are classified as dwarf galaxy candidates. Sources indicated by cross-hatched teal have magnitudes between -10.25 and -7.4, and are classified as GC candidates.}
    \label{fig:cut}
\end{figure}

We consider any ULX to be matched to an optical source if the latter is point-like and falls within the 95\% uncertainty region of the X-ray source added in quadrature with the inherent 0.8" uncertainty of the \textit{Chandra} pointing (as described in Section \ref{subsec:chandra}). We use TOPCAT\footnote{\url{http://www.star.bris.ac.uk/~mbt/topcat/}} to perform the matching. Of the 26 matches, 10 are double-matches, meaning 5 ULXs were each matched with 2 possible F110W counterparts within the X-ray uncertainty radius. The remaining 16 matches are single matches with each ULX source having a single F110W counterpart. 4 ULX candidates have no F110W counterpart, with 3 of them outside the range of the \textit{HST} observations.

To classify each optical match as a potential GC candidate or a background galaxy, we compare the absolute F110W magnitudes to theoretical magnitudes generated with stellar population synthesis models. We utilize a Python interface to the Flexible Stellar Population Synthesis library (Python-FSPS; \citealt{fsps1, fsps2}). This library requires a comprehensive set of parameters to initialize a new simulation, and outputs the magnitudes for a given \textit{HST} filter. Most of the parameters needed by Python-FSPS have little influence in the resulting flux measurements. We use the same set of parameters as those published by \cite{2017MNRAS.464..713P}, i.e., the Chabrier IMF and a GC age of 13 Gyr. We use an upper bound of $2\times10^6M_\odot$ and a lower bound of $2\times10^5M_\odot$ for the mass of a GC in the simulation. This results in a fainter threshold of $M_{F110W}$ = $-7.4$ and a brighter threshold of $M_{F110W}$ = $-10.25$ for GCs.

All the optical matches are brighter than the faint $M_{F110W}$ bound of $-7.4$. However, some sources are brighter than the $-10.25$ cutoff value. Ten of our sources fall within this range and therefore qualify as potential GCs. The sources above $M_{F110W}=-10.25$ are likely too bright to be GCs, but too dim to be regular galaxies, so we classify these as dwarf galaxy candidates. A histogram showing the magnitudes of the matches and the cutoff magnitude of $-10.25$ is shown in Figure \ref{fig:cut}.

\begin{table}
\caption{Number of background versus observed X-ray sources ($L_X\geq 7\times 10^{38}$ erg s$^{-1}$), calculated using methods discussed in Section \ref{sec:chandra}. The number of sources are shown for different galaxy effective radii.}
\begin{tabular}{lrrrrrr}
\hline\hline
Galaxy & N$_\mathrm{bkg}$ & N$_\mathrm{obs}$ & N$_\mathrm{bkg}$ & N$_\mathrm{obs}$ & N$_\mathrm{bkg}$ & N$_\mathrm{obs}$ \\
 & \multicolumn{2}{c}{(1 R$_\mathrm{eff}$)} & \multicolumn{2}{c}{(3 R$_\mathrm{eff}$)} & \multicolumn{2}{c}{(5 R$_\mathrm{eff}$)} \\ \hline
NGC-507 & 0.25 & 0 & 2.29 & 1 & 6.35 & 10 \\
NGC-708 & 0.27 & 1 & 2.39 & 8 & 6.64 & 16 \\
NGC-890 & 0.10 & 0 & 0.89 & 3 & 2.48 & 4 \\
NGC-1060 & 0.13 & 0 & 1.17 & 2 & 3.24 & 3 \\
NGC-1573 & 0.08 & 0 & 0.76 & 1 & 2.11 & 1 \\
NGC-1600 & 0.18 & 1 & 1.66 & 11 & 4.62 & 20 \\
NGC-1684 & 0.11 & 0 & 0.95 & 1 & 2.65 & 1 \\
NGC-1700 & 0.06 & 1 & 0.56 & 2 & 1.55 & 3 \\
NGC-5322 & 0.13 & 2 & 1.14 & 2 & 3.17 & 2 \\
NGC-5353 & 0.05 & 1 & 0.43 & 2 & 1.18 & 3 \\
NGC-5557 & 0.08 & 1 & 0.75 & 1 & 2.08 & 2 \\
NGC-6482 & 0.04 & 0 & 0.37 & 0 & 1.04 & 1 \\
NGC-7052 & 0.10 & 0 & 0.93 & 2 & 2.57 & 4 \\
NGC-7619 & 0.08 & 0 & 0.68 & 3 & 1.88 & 7 \\
NGC-7626 & 0.14 & 1 & 1.25 & 1 & 3.46 & 2 \\ \hline
Total & 1.80 & 8  & 16.21 &  40 & 45.03 & 79 \\ \hline
\end{tabular}
\label{tab:bkg} 
\end{table}

\section{Results}\label{sec:results}

\begin{table*}
    \caption{Final sample of GC ULX candidates with X-ray and optical measurements. A single X-ray source could match to 1 or 2 optical counterparts within the X-ray positional uncertainty (Column 4). The distance of the source from the galaxy centre in units of $R_\mathrm{eff}$ is shown in Column 5. For sources observed across multiple epochs, only the brightest $L_X$ (0.5 - 8 keV) is listed here (the full list is in Table \ref{tabA2:sources}).  The absolute magnitude and the classification for each optical counterpart is shown, although the coverage of F555W and F814W was limited and may not be available for all galaxies/optical matches. A classification of ``GC'' indicates that the optical counterpart was a globular cluster while ``DG'' indicates a likely dwarf galaxy.}
    \label{tab:sources}
    \centering
    \begin{tabular}{lccccccccc}
    \hline\hline
    Galaxy & ULX RA & ULX Dec & Err & \(R_\text{eff}\) & \(L_X\) & \(M_\text{F110W}\) & \(M_\text{F814W}\) & \(M_\text{F555W}\) & Classification \\
    & & & (arcsec) & & (10$^{39}$ erg s$^{-1}$) & & & & \\
    \hline
NGC-708 & 01:52:46.56 & +36:09:36.80    & 0.86 & 1.28 & 6.06$^{+1.03}_{-1.02}$ & -9.67 & --- & --- & GC \\
NGC-708 & 01:52:47.10 & +36:09:44.09 & 0.88 & 1.62 & 4.33$^{+1.81}_{-1.50}$ & -9.87 & --- & --- & GC \\
NGC-708 & 01:52:52.52 & +36:09:21.50 & 0.87 & 3.15 & 3.42$^{+0.82}_{-0.81}$ & -10.58/-12.42 & --- & --- & DG/DG \\
\hline
NGC-890 & 02:22:02.70 & +33:16:35.37 & 0.98 & 2.58 & 0.94$^{+0.82}_{-0.52}$ & -9.13 & --- & --- & GC \\
\hline
NGC-1060 & 02:43:18.41 & +32:25:15.43 & 1.13 & 2.68 & 3.16$^{+2.17}_{-1.60}$ & -9.66/-10.94 & --- & --- & GC/DG \\
\hline
NGC-1573 & 04:34:59.09 & +73:16:12.99 & 0.90 & 2.54 & 7.44$^{+3.80}_{-2.86}$ & -13.37 &--- & --- & DG \\
\hline
NGC-1600 & 04:31:39.83 & -05:05:52.36 & 0.35 & 2.04 & 4.29$^{+2.11}_{-1.63}$ & -10.99 & --- & --- & DG \\
NGC-1600 & 04:31:35.17 & -05:05:02.86 & 0.33 & 3.39 & 4.29$^{+3.07}_{-2.14}$ & -11.37 & --- & --- & DG \\
NGC-1600 & 04:31:37.34 & -05:04:57.20 & 0.48 & 1.92 & 1.60$^{+1.11}_{-0.77}$ & -10.23 & --- & --- & GC \\
NGC-1600 & 04:31:39.82 & -05:04:07.16 & 0.33 & 3.02 & 4.36$^{+2.96}_{-2.05}$ & -11.19 & --- &--- & DG \\
NGC-1600 & 04:31:40.14 & -05:05:40.90 & 0.38 & 1.50 & 3.36$^{+1.66}_{-1.28}$ & -10.26 &---  &--- & DG \\
NGC-1600 & 04:31:41.21 & -05:05:42.89 & 0.53 & 1.85 & 2.44$^{+1.21}_{-0.92}$ & -11.85 & --- & --- & DG \\
NGC-1600 & 04:31:42.53 & -05:04:36.40 & 0.66 & 2.50 & 1.74$^{+1.15}_{-0.82}$ & -9.64 & --- & --- & GC \\
NGC-1600 & 04:31:40.59 & -05:05:07.77 & 0.90 & 0.53 & 2.72$^{+1.72}_{-1.37}$ & -10.51/-13.28 & --- & --- & DG/DG \\
NGC-1600 & 04:31:41.52 & -05:04:20.95 & 0.95 & 2.64 & 1.05$^{+1.04}_{-0.69}$ & -11.04 & --- & --- & DG \\
\hline
NGC-7052 & 21:18:36.38 & +26:27:04.43 & 0.89 & 3.22 & 4.52$^{+2.00}_{-1.57}$ & -12.15/-9.38 & -11.10/--- &  --- & DG/GC \\
NGC-7052 & 21:18:33.77 & +26:27:22.99 & 0.98 & 2.42 & 1.40$^{+1.23}_{-0.77}$ & -10.47 & -10.08 & --- & DG \\
\hline
NGC-7619 & 23:20:11.90 & +08:11:25.11 & 0.64 & 4.70 & 6.45$^{+1.88}_{-1.58}$ & -10.67 & -10.63 & -9.61 & DG \\
NGC-7619 & 23:20:13.65 & +08:12:50.90 & 1.01 & 2.10 & 2.32$^{+1.36}_{-1.07}$ & -10.02 &-9.51 & -8.97 &  GC \\
NGC-7619 & 23:20:10.43 & +08:11:56.12 & 0.71 & 4.48 & 1.82$^{+1.26}_{-0.99}$ & -7.94 & --- & --- & GC \\
NGC-7619 & 23:20:14.40 & +08:11:25.75 & 1.06 & 3.85 & 1.20$^{+0.84}_{-0.64}$ & -9.76/-8.01 &-9.38/--- & --- &  GC/GC \\
    \hline
    \end{tabular}
\end{table*}

We find a total of 10 candidate GC ULXs across 21 galaxies (Table \ref{tab:sources}). Many of these galaxies contain bright X-ray gas within the central regions, which could effectively bury any ULXs that are located there. We do not include bright X-ray sources that were coincident with the centre of the galaxies in the final list of GC ULX candidates.

\subsection{Comparison to other GC ULXs}

\begin{figure}
    \centering
    \includegraphics[width=0.5\textwidth]{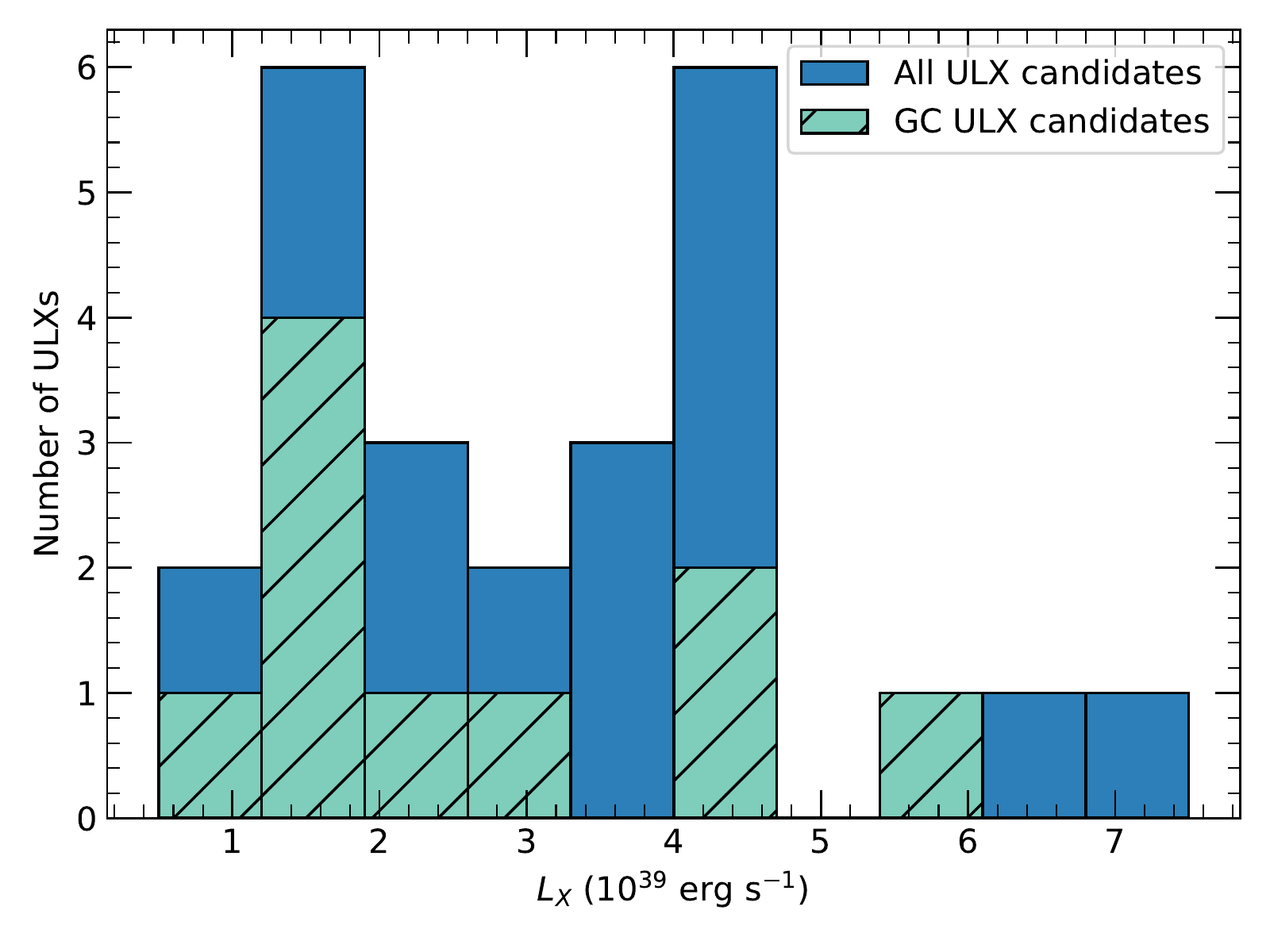}
    \caption{Number of ULXs per luminosity for $\leq 5 \times R_\mathrm{eff}$}
    \label{fig:lum}
\end{figure}

We find possible matches between X-ray sources and GCs out to the second-furthest galaxy in our sample (NGC 708, 69 Mpc). Due to the low signal-to-noise of the \textit{Chandra} observations, we cannot perform a systematic study of spectral or temporal properties of the candidate GC ULXs. These are important parameter spaces to explore for future studies to potentially uncover trends for GC ULXs. The only measurement we can reliably compare to previous GC ULX studies is the luminosities of these sources (Figure \ref{fig:lum}). We find a clustering around a few times $10^{39}$ erg s$^{-1}$, which is consistent with the range of luminosities found in previous searches \citep{2021MNRAS.504.1545D}. 
 
\subsection{Radial distance}

\begin{figure}
    \centering
    \includegraphics[width=0.5\textwidth]{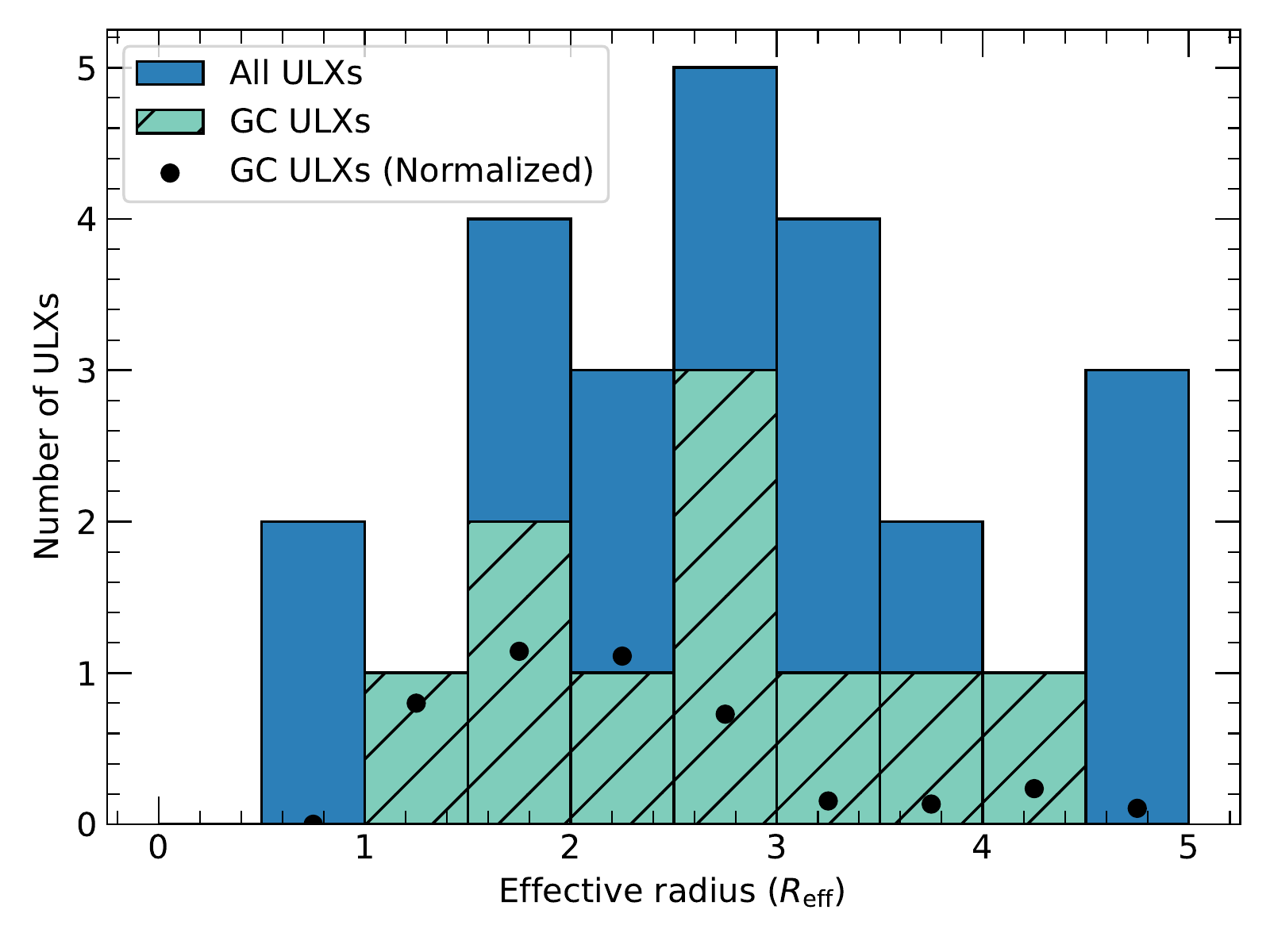}
    \caption{Number of ULXs per separation from host galaxy centre as a function of effective radius. The black dots show the number of GC ULXs normalized using the relative area at each effective radius. It is computed using intervals of $0.5\times R_\mathrm{eff}$ (step size of the histogram).}
    \label{fig:sep}
\end{figure}

The distribution of the entire sample of bright X-ray sources shows a peak at $\sim$3 $\times R_\mathrm{eff}$ (Figure \ref{fig:sep}). This is likely due to a combination of increased sensitivity close to the \textit{Chandra} aimpoint, combined with real galactic sources being more likely to exist in the inner regions of the galaxy. A significant decrease is apparent at $\lesssim 1 \times R_\mathrm{eff}$. Large quantities of hot X-ray gas at the central regions of the galaxies may also be possibly `burying' this population of ULXs.

The decrease in GC ULXs with increasing radius is likely influenced by several factors: there should inherently be fewer sources that are genuinely associated with the galaxy at further distances, and there can also be observational bias in the sense that the \textit{HST} images did not always cover the full $5 \times R_\mathrm{eff}$ regions. Given that the probability for contamination increases with distance from the galaxy core, it is difficult to determine whether or not the more distant sources are associated with the galaxy.

One avenue to place this work in context with the broader sample of GC ULXs is to compare the radial distributions of GC ULXs to previous studies. \cite{2021MNRAS.504.1545D} found that the majority of the 20 known GC ULXs reside within one effective radius of their host galaxy, with only one GC ULX located beyond this at $\sim8\times R_\mathrm{eff}$. This is different from the distribution of GC ULXs we identify shown in Figure \ref{fig:sep}, where most GC ULX candidates are within the range of $2\sim4.5 \times R_\mathrm{eff}$. However, it is not unexpected for GC ULXs to be located in the further regions of the host galaxies, as \cite{2020MNRAS.498.4790K} found a ``small but significant'' number of ULXs off-centre from elliptical galaxies which are possible GC ULXs. This suggests that even GC systems at the outskirts of galaxies may be host to peculiar X-ray sources, and it is worth follow-up in more nearby systems, where the outskirts are not often targeted.

\subsection{Host Galaxy Masses}

\begin{figure*}
    \centering
    \includegraphics[width=0.9\textwidth]{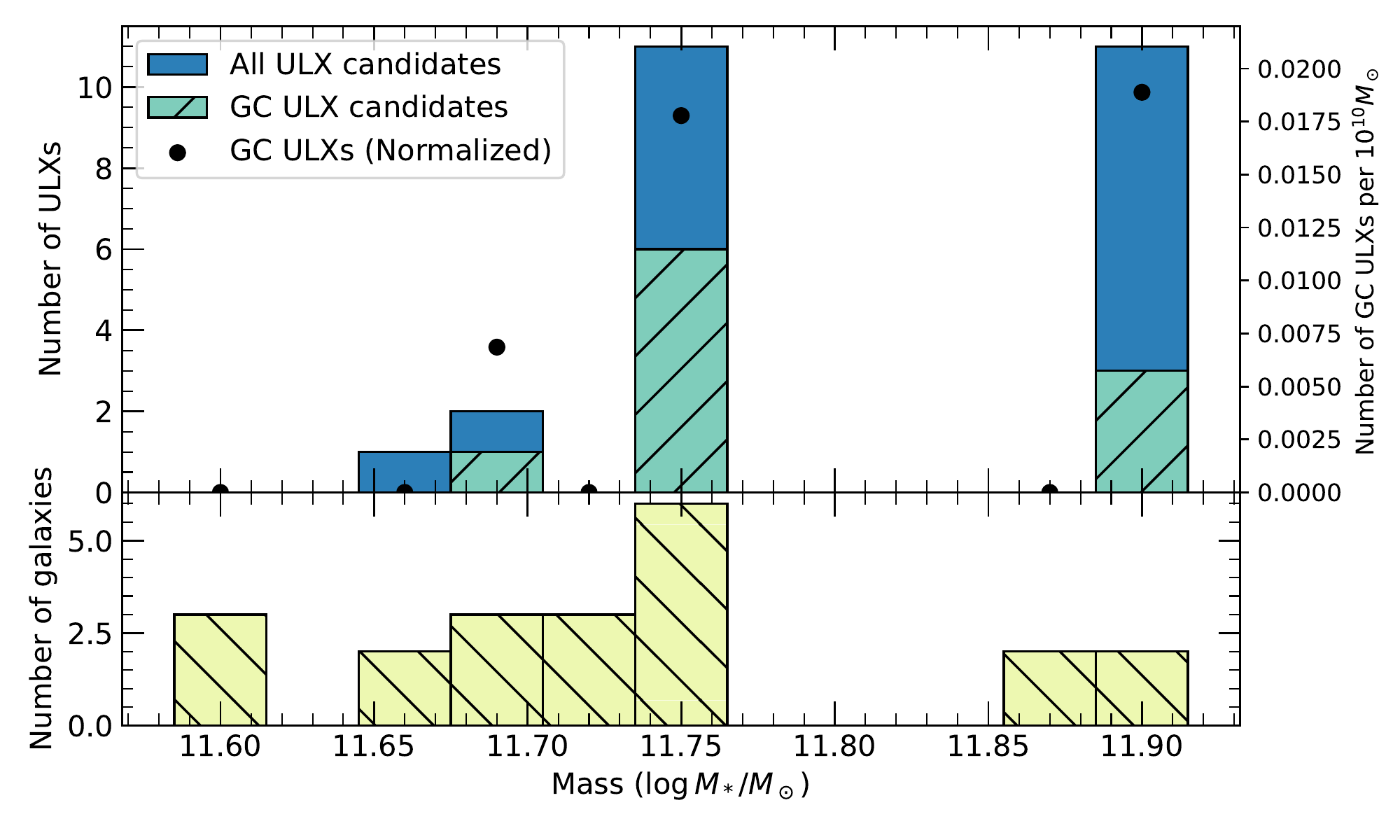}
    \caption{The upper panel shows the number of total ULXs and GC ULX candidates for each of the 21 galaxies as a function of the galaxies' mass. The black dots show the number of GC ULXs normalized using the sum of the galaxies' mass. It is computed using intervals of $0.03\times(\log{M_*/M_\odot})$ (step size of the histogram). The lower panel shows a histogram of the galaxies' mass in our sample.}
    \label{fig:mass}
\end{figure*}

The distribution of the GC ULXs as a function of the host galaxies' mass is shown in the upper panel of Figure \ref{fig:mass}, while the number of analysed galaxies at each mass is shown in the lower panel to demonstrate the range of masses in the sample. We also compute the number of GC ULXs normalized using the total mass of the galaxies with sufficient data. We find four distinct mass ranges of the sample galaxies:

\begin{enumerate}
    \item For galaxies with mass near $11.60\times(\log{M_*/M_\odot})$, we find no GC ULXs in our sample. We note that two of the three galaxies in this mass range (NGC 128 and 1066) do not have sufficient \textit{HST} coverage (see Section \ref{sec:noHST}).
    \item Out of the 5 galaxies with mass near $11.68\times(\log{M_*/M_\odot})$, only one (NGC 890) contains a single GC ULX, which gives a ``GC ULX rate'' of $\sim0.2$ GC ULXs per galaxy and $\sim0.004$ GC ULXs per $10^{10}M_\odot$. All 5 galaxies in this mass range have sufficient data for analysis. 
    \item A total of nine galaxies have mass near $11.75\times(\log{M_*/M_\odot})$, with six GC ULXs found across multiple host galaxies. This is the dominant mass range where the analysis is carried out. All galaxies in this mass range have sufficient optical and X-ray data and we conclude a ``GC ULX rate'' of $\sim0.67$ GC ULXs per galaxy and $\sim0.012$ GC ULXs per $10^{10}M_\odot$.
    \item Four galaxies have mass near $11.87$ ($\log{M_*/M_\odot}$) with a total of three GC ULX candidates - two of these residing in NGC 1600. Of the four galaxies in this mass range, three have sufficient optical data for analysis, and we estimate a ``GC ULX rate'' of $\sim1.0$ GC ULX per galaxy and $\sim0.013$ GC ULXs per $10^{10}M_\odot$.
\end{enumerate}

The computed ``GC ULX rate'' and the number of GC ULXs per mass tentatively show an increase with galaxy mass, however, further studies will need to confirm this trend.

We also calculate the number of GC ULXs per unit stellar mass using all GC ULXs identified here, excluding the three galaxies mentioned in Section \ref{sec:sample_selection} that have insufficient \textit{HST} coverage. Using the masses of the remaining 18 galaxies, we compute the number of GC ULXs per unit stellar mass of $0.010$ per $10^{10}M_\odot$ which corresponds to 1 GC ULX per $10^{12}M_\odot$ of galaxy stellar mass.

To determine the rate of GC ULXs per number of sources detected with $L_X \gtrsim7\times10^{38}$ erg s$^{-1}$, we compute the ratio of GC ULXs at each mass range to the detected X-ray sources with $L_X \gtrsim7\times10^{38}$ erg s$^{-1}$ (79 in total). We obtain $0.0$, $0.083$, $0.182$ and $0.125$ for the same four mass ranges defined previously, from low to high galaxy mass. Overall, for the entire sample, this rate is $0.127$ GC ULXs per all ULXs detected.

\subsection{ULX recovery in hot gas}

We repeat our analysis process for the well-studied, nearby galaxy M87 (which was part of the original MASSIVE Survey; see Section \ref{sec:sample_selection}) to test the completeness of our identification of ULXs within the central regions of the galaxies shrouded in hot gas. M87 is a nearby (16.8 Mpc) analogue of the galaxies in our study, as it is an elliptical galaxy with significant hot X-ray gas, and has seven candidate GC ULXs \citep{2020MNRAS.497..596D}, three of which are located in the varying background of the hot gas. While the GC ULX population of M87 has been studied extensively, previous work has focused on the inner regions of the galaxy. The inner regions of galaxies become harder to study at large distances because the luminosity of the background gas in a given angular resolution element increases while the luminosities of point sources do not change. This provides an opportunity to compare the detection of ULXs within hot gas from our sample to a much closer analogue with higher signal-to-noise, and in which the hot gas has distinguishable features.

We reduce the deepest image of M87 (ObsID 2707, ACIS-S, exposure time 98.66 ks) using the same method described in \ref{subsec:chandra}, and bin the image to 2 and 4 times the original pixel scale, effectively placing the galaxy at $\sim33.6$ Mpc and $\sim 67.2$ Mpc in terms of observed spatial scale. These binned pixel scales approximately correspond to those of the nearest and farthest galaxies in our sample. We run the {\tt wavdetect} algorithm on the binned and unbinned images to compare the recovery of six GC ULXs in M87 (L18-GCULX, J04-GCULX, M87-GCULX1, M87-GCULX2, M87-GCULX3 and M87-GCULX4 from Table 2 of \citealt{2020MNRAS.497..596D}). All ULXs are successfully detected by {\tt wavdetect} in the unbinned images. L18-GCULX, M87-GCULX1 and M87-GCULX4 are detected in both binned images, and are the furthest of these sources from the galaxy centre. M87-GCULX3 is recovered in the image binned by a factor of two, but is blended with a second nearby source in the factor of four binning. This indicates that source blending may increase the luminosity of some ULXs at these large distances, though it is not likely to occur for all sources. M87-GCULX2 is detected in both binned images, but in the image binned by a factor of four has a large {\tt wavdetect} uncertainty region and peaks in a single pixel. Finally, the most central source, J04-GCULX, is detected in the factor of two binning but is not recovered at all in the larger binning, effectively becoming indiscernible from the hot gas. This supports the idea that we are unlikely able to detect ULXs buried in the central hot gas within our sample. Note that only one galaxy in our sample, NGC-708, has an observation with comparable exposure time, but is also a factor of four further away, meaning the signal-to-noise would be significantly lower for all of the galaxies in this study.

\subsection{Comparison of sources}

Due to the large distances of the galaxies and relatively short exposure times of the \textit{Chandra} observations, we are not able to extract spectra for the GC ULX candidates as most have low net counts. We instead calculate the hardness ratio of the sources in each epoch (see Table \ref{tabA2:sources}) based on the observed net counts for each source as a rudimentary way to analyse the spectral information of these sources. Here we define the hardness ratio as $(s-h)/(s+h)$, where $s$ and $h$ are the net source counts in the soft (0.5-2 keV) and hard (2-8 keV) energy bands, respectively. Thus, a `hard' source will have a negative ratio, and a `soft' source will be positive.

We run a series of 2-sample Anderson-Darling tests to quantify the independence of the hardness ratio distributions for sources we classify as `globular clusters' versus `dwarf galaxies'. Two of our X-ray sources each have two possible optical counterparts with different classifications (within NGC 1060 and NGC 7052; see Table \ref{tab:sources}), so we run each set of tests four times to include all combinations of classifications. In the first test, we use the hardness ratios across all epochs for each source, i.e., sources with multiple \textit{Chandra} observations are counted multiple times. In all four scenarios, the resulting p-value is $>0.25$, indicating that we cannot prove the hardness ratios are drawn from different distributions. For the second test, we calculate the average hardness ratio weighted by the inverse variance for sources with multiple observations, so they are only included once in the sample, and run the Anderson-Darling tests again including all four scenarios. In this case, all but one combination has a p-value $>0.23$. The null hypothesis of the distributions being drawn from the same sample can be rejected at marginal level (p$=0.054$) only when the optical counterpart to the ULX in NGC 1060 is classified as a GC and that of the more luminous ULX in NGC 7052 is classified as a DG. As this is a limited sample size of sources with low counts, and due to the contrived nature of the one scenario in which we can reject the null hypothesis, we do not interpret these results as being useful to understanding any underlying properties to help classify ULX associations.

\section{Galaxy Summaries}
\label{sec:gal_summaries}
We examine a total of 21 early-type galaxies, obtained by limiting our sub-sample of the MASSIVE survey \citep{2014ApJ...795..158M} to $\leq70$ Mpc and excluding galaxies that have been previously studied for their GC ULX populations. A few galaxies have significant contamination from X-ray gas. Some X-ray sources are erroneously detected because of this, so we exclude these sources from our final list. We also exclude any X-ray source that could be the central super-massive black hole.

\subsection{Galaxies with new GC ULXs}

\subsubsection{NGC 708 $-$ 2 GC ULXs}

NGC 708 has 16 detected ULXs, with 14 of these observed in the \textit{HST} field of view (FOV). Six do not have optical counterparts. Five other sources appear to be associated with other galaxies, with one of these coincident with the outskirts of a faint galaxy only visible in the F110W filter. Three sources are coincident with optical point sources: two were classified as GCs, and one as a DG. This is one of the most distant galaxies in our sample ($\sim$69 Mpc), showing that it is feasible to detect ULXs with optical associations within very distant galaxies.

\subsubsection{NGC 890 $-$ 1 GC ULX}

NGC 890 has some central X-ray emission. There are four detected ULXs - one has an \textit{HST} counterpart, two have no detected \textit{HST} counterparts, and one source is not in the \textit{HST} FOV. The ULX candidate with an optical counterpart has a luminosity of $L_\mathrm{X} = 9.4 \times 10^{38}$ erg/s, but the upper limit is $1.8 \times 10^{39}$ erg/s, and we classify the host as a GC.

\subsubsection{NGC 1060 $-$ 1 GC ULX}

There are three ULXs candidates, but the \textit{Chandra} images are very low signal-to-noise. For one of the sources, the positional uncertainty does not clearly coincide with a \textit{HST} counterpart. One ULX candidate has a somewhat extended counterpart which is likely a galaxy, and the third ULX candidate has two F110W matches, one of which we classify as a GC and the other as a DG. 

\subsubsection{NGC 1600 $-$ 2 GC ULXs}

There are 20 X-ray sources above the ULX luminosity threshold, but five of these are not in the FOV of \textit{HST}. Of the 10 that match to counterparts in \textit{HST}, we classify eight as DGs and only two as GCs.

\subsubsection{NGC 7052 $-$ 1 GC ULX}

NGC 7052 has strong central diffuse emission and four off-nuclear sources. One source  appears to be coincident with another galaxy and another has no optical counterpart. Two sources coincide with point-like optical sources. We classify one of these counterparts as a GC and the other as a DG.

\subsubsection{NGC 7619 $-$ 3 GC ULXs}

NGC 7619 has strong central emission, and seven off-nuclear sources. Three sources each have a point-like optical counterpart which we classify as GCs, and one is coincident with a slightly extended optical source we classify as a DG. The remaining three sources are excluded from our study as they either do not have a detected optical counterpart or a potential counterpart is at the edge of the X-ray uncertainty region.

Optical counterparts for two ULXs in NGC 7619 are identified in both F555W ($V$ filter) and F814W ($I$ filter; see Table \ref{tab:sources}).  The optical source we classify as a GC has an optical colour of $V-I=0.54$, indicating that it is metal-poor, while the one we classify as a DG has a colour of $V-I=1.02$.

\subsection{Galaxies with no new GC ULXs}

NGC 507 contains 10 X-ray sources with  $L_\mathrm{X} \geq 7 \times 10^{38}$ erg s$^{-1}$, with only four of these falling within at least one \textit{HST} image. Two of the sources appear to be central emission from other galaxies and one has no detected optical counterpart, leaving one ULX candidate that is coincident with a point source detected with the F110W filter. However, this source is extended in X-ray and therefore unlikely to be a ULX. NGC 507 has strong central X-ray emission from hot gas, so it is possible that more ULXs are hidden closer to the centre of the galaxy. 

NGC 1453 has no off-nuclear X-ray sources. One ULX was detected in NGC 1573, and it has a large, point-like optical counterpart which we classify as a dwarf galaxy. NGC 1684 has some central emission and one off-nuclear source. The X-ray source is coincident with a spatially extended optical counterpart which is likely a galaxy. NGC 1700 has very strong central emission and three X-ray sources above the ULX limit. One is coincident with a star, and the remaining two sources do not have \textit{HST} counterparts. NGC 2258 and NGC 2672 have central diffuse emission but no off-nuclear X-ray sources. NGC 5322 has strong central X-ray emission and two bright X-ray sources - neither of which have detected optical counterparts.

NGC 5353 has strong central X-ray emission and three ULXs that do not have detected counterparts. NGC 5557 has central diffuse emission and two ULXs. One source has a very extended counterpart, likely a galaxy. The second source has no detected counterparts. NGC 6482 has extremely strong central X-ray emission. It has one ULX with no detected optical counterpart. NGC 7626 has central X-ray emission and two bright X-ray sources. One of the sources does not have an optical counterpart, and the second has a counterpart which is at the edge of the X-ray uncertainty and is thus not included in the final sample.

\subsection{Galaxies with insufficient data}
\label{sec:noHST}
NGC 128 has one central source, and one off-nuclear source that falls outside of the \textit{HST} FOV. The only \textit{HST} images available were obtained using the NIC3 camera, which is not sufficient for matching with the X-ray sources. As there was no way to determine any optical counterpart for the one off-nuclear source, we do not include this in the full investigation. NGC 499 and NGC 1066 both have strong central diffuse emission, and host one and five off-nuclear sources, respectively. However, there were no archival \textit{HST} data for either galaxy at the time of this study, so we exclude these from our analysis.

\section{Discussion}
\label{section:discusson}
We analysed archival \textit{Chandra} X-ray and \textit{HST} infrared/optical observations of 21 external galaxies to search for GC ULXs. Of these 21, only 15 had \textit{HST} observations at the time of study, with F110W as the uniting filter. We implemented flexible stellar synthesis population modeling to classify the IR sources as either globular clusters or dwarf galaxies. 

This study represents the most spatially complete study of GC ULXs in a given galaxy. Past studies of more nearby galaxies, e.g. \cite{2020MNRAS.497..596D}, have been limited to searching smaller regions of the galaxies with already existing \textit{Chandra} and \textit{HST} data. However, due to the relative projection of sizes on the sky because of the larger distances, many of the galaxies in this study had \textit{Chandra} observations which covered at least 5 $\times R_{\textrm{eff}}$, with mostly complete \textit{HST} coverage. This enables us for the first time to place mostly complete constraints on the rates of GC ULXs. Such rates will have significant implications for comparison to theoretical work on understanding the rates of black holes in GCs (e.g. \citealt{askar17,Leveque22, Leveque2022a}). Studies such as the MOCCA-survey database (e.g. \citealt{Leveque22,Leveque2022a}) are merging population synthesis codes with star cluster evolution code to place theoretical constraints on the rates of black holes in globular clusters. Although current results are focused on modeling globular cluster populations in spiral galaxies, an initial comparison of observable properties between \cite{Leveque22} and this study is promising. Notably, \cite{Leveque22}'s results suggest that globular clusters that harbour black holes are more massive (more optically luminous), which we also see some evidence for in this sample, along with the sample in \citet{2021MNRAS.504.1545D}. \cite{Leveque22} also find that globular clusters in the galactic outskirts have a higher probability of hosting a massive black hole, or black hole system. In our study, we found that most of the GC ULXs are within 2 effective radii, suggesting that there are either differences in the probabilistic model trends for spiral galaxies, rather than elliptical, or that many of the GC ULXs are not markers of massive black holes, but more so stellar mass compact objects.

While the primary focus of this study is on the association of ULXs with globular clusters, a number of the X-ray sources we uncovered are hosted by dwarf galaxies. The study of black holes in dwarf galaxies has a number of exciting astrophysical implications to understand the formation of supermassive black holes and onset of star formation, e.g., \cite{Cann2021, Schutte}, among many others. The sources we identify, along with the properties of the host galaxies, may be useful in helping constrain the low-mass nuclear black hole function.

Our main results are summarised as follows: 
\begin{itemize}
    \item Although our study is biased towards detecting the brightest optical counterparts due to the large distances of the host galaxies, we find a maximum of 10 GC ULX candidates.
    \item We have identified GC ULX candidates out to a distance of 69 Mpc (NGC 708) in the archival data.
    \item We compare the spatial and X-ray luminosity distributions of this sample to previously studied GC ULXs and find that both populations tend to be within 2 effective radii of their host galaxies and peak at a few $\times 10^{39}$ erg/s. 
    \item We do not find a clear trend between the mass of the host galaxy and the number of GC ULXs, however, we are limited by the availability of archival data and do not probe a sufficiently dynamic range of galaxy masses to make a strong claim. 
    \item As most X-ray sources studied here do not have a sufficient number of counts, we cannot investigate spectral properties of these sources. Instead, we analyse the distribution of hardness ratios of the sources, but find no evidence that the DG and GC optical counterparts preferentially host ULXs with different hardness ratios.
\end{itemize}
As the brightest X-ray emitters in old, dynamic globular clusters, GC ULXs may unlock some of the secrets pertaining to peculiar high energy astrophysical events in globular clusters. We recommend future studies to expand the optical coverage of these galaxies and to better identify globular cluster counterparts, as well as more extensive X-ray searches to identify these unique sources. 

\section*{Acknowledgements}
The authors thank the anonymous referee for their insightful comments on this paper. YS, KCD and DH acknowledge funding from the Natural Sciences and Engineering Research Council of Canada (NSERC), the Canada Research Chairs (CRC) program. KCD acknowledges fellowship funding from the McGill Space Institute and from Fonds de Recherche du Qu\'ebec $-$ Nature et Technologies, Bourses de recherche postdoctorale B3X no. 319864. This work was performed in part at Aspen Center for Physics, which is supported by National Science Foundation grant PHY-1607611.

Based on observations made with the NASA/ESA Hubble Space Telescope, and obtained from the Hubble Legacy Archive, which is a collaboration between the Space Telescope Science Institute (STScI/NASA), the Space Telescope European Coordinating Facility (ST-ECF/ESAC/ESA) and the Canadian Astronomy Data Centre (CADC/NRC/CSA).
\section*{Data Availability}
The \textit{Chandra} observations are publicly available at \url{https://cda.harvard.edu/chaser/}. The \textit{HST} data are publicly available through the Mikulski Archive for Space Telescopes \url{https://mast.stsci.edu/portal/Mashup/Clients/Mast/Portal.html}.

\setlength{\bibsep}{0pt}
\bibliographystyle{mnras}
\bibliography{references}

\appendix

\section{Additional Tables}
The archival \textit{Chandra} and \textit{HST} observations available for all the galaxies in our sample is shown in Table \ref{tab:obs}.

\clearpage
\newpage
\begin{table*}
\caption{Archival \textit{Chandra} and \textit{HST} observations of the galaxies in this sample.}
\scriptsize
\begin{tabular}{lllll|lllll}

\hline \hline
Galaxy   & ObsID     & \multicolumn{1}{c}{Instrument} & Exp  & Date       & Prop ID & Detector & Filter & Exp & Date  \\
  & (\textit{Chandra}) &    & (ks) &     &  (HST)      &   & & (s) &       \\ \hline
NGC-499, NGC-507  & 317       & ACIS-S    & 26.85     & 2000-10-11 & 6587   & WFPC2    & F555W  & 1700     & 1997-07-13   \\
  & 2882      & ACIS-I    & 43.63     & 2002-01-08 & 14219  & WFC3/IR  & F110W  & 2496.17  & 2015-11-20   \\
  & 10536     & ACIS-S    & 18.39     & 2009-02-12 &        &   & &   &       \\
  & 10865     & ACIS-S    & 5.11      & 2009-02-04 &        &   & &   &       \\
  & 10866     & ACIS-S    & 8.04      & 2009-02-05 &        &   & &   &       \\
  & 10867     & ACIS-S    & 7.06      & 2009-02-07 &        &   & &   &       \\ \hline
NGC-708  & 2215      & ACIS-S    & 28.74     & 2001-08-03 & 5910   & WFPC2    & F814W  & 16400    & 1996-02-11   \\
  & 7921      & ACIS-S    & 110.67    & 2006-11-20 & 7281   & WFPC2    & F555W  & 300      & 1999-09-01   \\
  &    &    &    &     & 7281   & WFPC2    & F814W  & 300      & 1999-09-01   \\
  &    &    &    &     & 14219  & WFC3/IR  & F110W  & 2496.17  & 2016-01-02   \\ \hline
NGC-890  & 18035     & ACIS-S    & 5.97   & 2015-11-27 & 14219  & WFC3/IR  & F110W  & 2496.17  & 2015-11-28   \\
  & 19325     & ACIS-S    & 34.6   & 2016-10-03 &        &   & &   &       \\ \hline
NGC-1060, NGC-1066 & 18269     & ACIS-I    &  24.17  & 2015-11-18 & 14219  & WFC3/IR  & F110W  & 2496.17  & 2015-11-22   \\
  & 18712     & ACIS-I    &  21.78  & 2015-12-22 &        &   & &   &       \\
  & 18713     & ACIS-I    &  29.19  & 2015-12-12 &        &   & &   &       \\
  & 18714     & ACIS-I    &  14.88  & 2015-11-26 &        &   & &   &       \\
  & 18715     & ACIS-I    &  21.79  & 2015-11-20 &        &   & &   &       \\
  & 18716     & ACIS-I    & 14.85   & 2015-12-24 &        &   & &   &       \\
  & 19311     & ACIS-S    & 9.83    & 2016-12-11 &        &   & &   &       \\
  & 19312     & ACIS-S    & 16.81   & 2016-11-22 &        &   & &   &       \\ \hline
NGC-1453 & 19314    & ACIS-S    & 7.93    & 2016-11-23 &  14219 & WFC3/IR & F110W & 2496.17 & 2016-08-19 \\ \hline
NGC-1573 & 19315     & ACIS-S    & 13.78   & 2015-11-29 & 14219  & WFC3/IR  & F110W  & 2895.4   & 2015-11-29   \\ \hline
NGC-1600 & 4283      & ACIS-S    & 26.78   & 2002-09-18 & 14219  & WFC3/IR  & F110W  & 2496.17  & 2016-09-01   \\
  & 4371      & ACIS-S    & 26.75   & 2002-09-20 &        &   & &   &       \\
  & 21374     & ACIS-S    & 25.72   & 2018-12-03 &        &   & &   &       \\
  & 21375     & ACIS-S    & 42.21   & 2019-11-28 &        &   & &   &       \\
  & 21998     & ACIS-S    & 13.87   & 2018-12-03 &        &   & &   &       \\
  & 22878     & ACIS-S    & 44.97   & 2019-11-25 &        &   & &   &       \\
  & 22911     & ACIS-S    & 31.01   & 2019-11-01 &        &   & &   &       \\
  & 22912     & ACIS-S    & 35.64   & 2019-11-02 &        &   & &   &       \\ \hline
NGC-1684 & 19316     & ACIS-S    &  15.73  & 2016-11-22 & 14219  & WFC3/IR  & F110W  & 2496.17  & 2016-09-04   \\ \hline
NGC-1700 & 2069      & ACIS-S    &  42.8  & 2000-11-03 & 5416   & WFPC2    & F555W  & 1660     & 1994-09-08   \\
  &    &    &    &     & 5416   & WFPC2    & F814W  & 1260     & 1994-09-09   \\
  &    &    &    &     & 5454   & WFPC2    & F555W  & 1000     & 1994-04-17   \\
  &    &    &    &     & 5454   & WFPC2    & F814W  & 460      & 1994-04-17   \\
  &    &    &    &     & 6587   & WFPC2    & F555W  & 1600     & 1997-08-08   \\
  &    &    &    &     & 6587   & WFPC2    & F814W  & 1800     & 1997-08-08   \\
  &    &    &    &     & 14219  & WFC3/IR  & F110W  & 2496.17  & 2016-09-12   \\ \hline
 NGC-2258 & 19317 & ACIS-S & 7.66 & 2017-05-06 & 8212 & WFPC2 & F814W & 13500 & 2000-02-15 \\
          &       &        &      &            & 8212 & WFPC2 & F814W & 2700 & 2000-02-17 \\ \hline
 NGC-2672 & 19319 & ACIS-S & 11.31 & 2017-02-17 &  14219   & WFC3/IR  &  F110W &  2496.17 & 2016-04-06 \\ \hline
NGC-5322 & 6787      & ACIS-S    & 13.77   & 2006-08-20 & 5454   & WFPC2    & F555W  & 1000     & 1994-11-26   \\
  &    &    &    &     & 5454   & WFPC2    & F814W  & 460      & 1994-11-26   \\
  &    &    &    &     & 9427   & ACS/WFC  & F814W  & 820      & 2002-07-25   \\
  &    &    &    &     & 14219  & WFC3/IR  & F110W  & 2745.4   & 2015-11-17   \\ \hline
NGC-5353 & 5903      & ACIS-I    &  4.49  & 2005-04-10 & 14219  & WFC3/IR  & F110W  & 2496.17  & 2015-12-18 \\
  & 14903     & ACIS-S    & 40.27   & 2014-03-31 & & & & & \\ \hline
NGC-5557 & 19324     & ACIS-S    &  8.83  & 2016-12-17 & 6587   & WFPC2    & F555W  & 1400     & 1997-06-03   \\
  &    &    &    &     & 9427   & ACS/WFC  & F814W  & 2400     & 2002-09-03   \\
  &    &    &    &     & 14219  & WFC3/IR  & F110W  & 2496.17  & 2016-03-20   \\ \hline
NGC-6482 & 3218      & ACIS-S    &   19.35 & 2002-05-20 & 14219  & WFC3/IR  & F110W  & 2496.17  & 2016-06-11   \\
  & 19584     & ACIS-S    & 27.61   & 2017-11-14 &   & & & &   \\
  & 19585     & ACIS-S    & 19.7   & 2018-02-22 &        &   & &   &       \\
  & 20850     & ACIS-S    & 19.7   & 2017-11-24 &        &   & &   &       \\
  & 20857     & ACIS-S    & 22.95   & 2017-11-26 &        &   & &   &       \\
  & 20978     & ACIS-S    & 19.88   & 2018-02-22 &        &   & &   &       \\
  & 20979     & ACIS-S    &  9.82  & 2018-02-23 &        &   & &   &       \\
  & 20980     & ACIS-S    &  75.66  & 2018-05-29 &        &   & &   &       \\ \hline
NGC-7052 & 2931      & ACIS-S    &  9.63  & 2002-09-21 & 5848   & WFPC2    & F814W  & 1470     & 1995-06-23   \\
  & 19326     & ACIS-S    & 38.55   & 2017-09-30 & 7281   & WFPC2    & F555W  & 700      & 1999-04-16   \\
  &    &    &    &     & 14219  & WFC3/IR  & F110W  & 2496.17  & 2016-06-14   \\ \hline
NGC-7619 & 2074      & ACIS-I    &  26.74  & 2001-08-20 & 6554   & WFPC2    & F555W  & 2200     & 1998-10-10   \\
  & 3955      & ACIS-S    &  37.47  & 2003-09-24 & 6554   & WFPC2    & F814W  & 2200     & 1998-10-10   \\
  &    &    &    &     & 9427   & WFPC2    & F814W  & 4200     & 2003-07-27   \\
  &    &    &    &     & 14219  & WFC3/IR  & F110W  & 2496.17  & 2016-06-17   \\ \hline
NGC-7626 & 2074      & ACIS-I    & 26.74   & 2001-08-20 & 5454   & WFPC2    & F555W  & 1000     & 1994-05-28 \\
  & 3955      & ACIS-S    & 37.47   & 2003-09-24 & 5454   & WFPC2    & F814W  & 460      & 1994-05-29  \\
  &    &    &    &     & 9399   & ACS/WFC  & F814W  & 960      & 2002-10-07 \\
  &    &    &    &     & 9427   & ACS/WFC  & F814W  & 2600     & 2003-07-27
\end{tabular}
\label{tab:obs} 
\end{table*}

\begin{table*}
\scriptsize
    \caption{Measured counts and fluxes from final sample of GC ULX candidates in the 0.5-8.0 keV band. The hardness ratios are calculated as $(s-h)/(s+h)$, where $s$ and $h$ are the net counts in the soft (0.5-2 keV) and hard (2-8 keV) bands, respectively. A classification of ``GC'' indicates that the optical counterpart was a globular cluster while ``DG'' indicates a likely dwarf galaxy. The flux and luminosity are the model-corrected unabsorbed values, while the net counts and count rate are the observed background-corrected counts. All measurements are using the 0.5-8 keV band. }
    \label{tabA2:sources}
    \centering
\begin{tabular}{lllrrrccrc}
\hline\hline
Galaxy & ULX RA & ULX Dec & ObsID & Net Counts & Net Count Rate & Flux & $L_X$ & Hardness Ratio & Classification \\
 &  &  &(\textit{Chandra}) &  & ($10^{-4} \rm{s}^{-1}$) & ($10^{-15}$ erg s$^{-1}$ cm$^{-2}$) & ($10^{39}$ erg s$^{-1}$ cm$^{-2}$) &  &  \\ \hline
NGC-0708 & 01:52:46.56 & +36:09:36.80 & 7921 & $115.00\pm11.84$ & $10.41\pm1.07$ & $10.60_{-1.76}^{+1.80}$ & $6.06^{+1.03}_{-1.02}$ & $0.37\pm0.09$ & GC \\
\rowcolor{lightgray}NGC-0708 & 01:52:47.10 & +36:09:44.09 & 2215 & $25.42\pm5.84$ & $9.00\pm2.07$ & $7.60_{-2.63}^{+3.20}$ & $4.33_{-1.50}^{+1.81}$ & $0.22\pm0.22$ & GC \\
NGC-0708 & 01:52:52.52 & +36:09:21.50 & 7921 & $72.09\pm10.41$ & $6.53\pm0.94$ & $6.01_{-1.43}^{+1.43}$ & $3.42_{-0.81}^{+0.82}$ & $0.58\pm0.11$ & DG/DG \\ \hline
NGC-0890 & 02:22:02.70 & +33:16:35.37 & 19325 & $6.00\pm2.45$ & $1.76\pm0.72$ & $2.55_{-1.42}^{+2.23}$ & $0.94_{-0.52}^{+0.82}$ & $0.00\pm0.41$ & GC \\\hline
NGC-1060 & 02:43:18.41 & +32:25:15.43 & 18713 & $9.90\pm3.50$ & $3.39\pm1.20$ & $5.82_{-2.95}^{+3.98}$ & $3.16_{-1.60}^{+2.17}$ & $-0.46\pm0.34$ & GC/DG \\\hline
NGC-1573 & 04:34:59.09 & +73:16:12.99 & 19315 & $14.58\pm3.88$ & $10.60\pm2.82$ & $14.70_{-5.64}^{+7.50}$ & $7.44_{-2.86}^{+3.80}$ & $0.37\pm0.25$ & DG \\\hline
NGC-1600 & 04:31:39.83 & -05:05:52.36 & 4283 & $21.04\pm4.70$ & $9.38\pm2.10$ & $7.91_{-2.60}^{+3.29}$ & $3.85_{-1.27}^{+1.61}$ & $0.54\pm0.19$ & DG \\
 &  &  & 4371 & $27.08\pm5.51$ & $10.12\pm2.06$ & $8.39_{-2.56}^{+3.11}$ & $4.09_{-1.24}^{+1.52}$ & $0.45\pm0.19$ &  \\
 &  &  & 21374 & $15.94\pm4.14$ & $6.20\pm1.61$ & $8.80_{-3.34}^{+4.30}$ & $4.29_{-1.63}^{+2.11}$ & $0.06\pm0.19$ &  \\
 &  &  & 21375 & $10.09\pm3.34$ & $2.39\pm0.79$ & $3.84_{-1.80}^{+2.52}$ & $1.87_{-0.88}^{+1.22}$ & $0.10\pm0.33$ &  \\
 &  &  & 22878 & $14.99\pm4.02$ & $3.33\pm0.89$ & $5.12_{-1.99}^{+2.61}$ & $2.49_{-0.97}^{+1.27}$ & $-0.12\pm0.27$ &  \\
 &  &  & 22911 & $13.31\pm3.75$ & $4.33\pm1.22$ & $6.58_{-2.68}^{+3.62}$ & $3.20_{-1.31}^{+1.74}$ & $0.33\pm0.27$ &  \\
 &  &  & 22912 & $13.27\pm3.76$ & $3.73\pm1.06$ & $6.09_{-2.49}^{+3.32}$ & $2.96_{-1.21}^{+1.62}$ & $-0.33\pm0.27$ &  \\
\rowcolor{lightgray}NGC-1600 & 04:31:35.17 & -05:05:02.86 & 4283 & $16.45\pm4.13$ & $4.33\pm1.84$ & $6.18_{-2.25}^{+2.96}$ & $3.01_{-1.10}^{+1.44}$ & $0.65\pm0.19$ & DG \\
\rowcolor{lightgray} &  &  & 4371 & $27.31\pm5.30$ & $10.21\pm1.98$ & $8.53_{-2.47}^{+3.07}$ & $4.16_{-1.21}^{+1.47}$ & $0.58\pm0.16$ &  \\
\rowcolor{lightgray} &  &  & 21374 & $7.57\pm2.84$ & $2.94\pm1.10$ & $4.56_{-2.37}^{+3.48}$ & $2.22_{-1.15}^{+1.70}$ & $0.81\pm0.25$ &  \\
\rowcolor{lightgray} &  &  & 21375 & $23.00\pm5.03$ & $5.45\pm1.19$ & $8.13_{-2.65}^{+3.27}$ & $3.99_{-1.29}^{+1.60}$ & $0.08\pm0.22$ &  \\
\rowcolor{lightgray}&  &  & 22878 & $17.56\pm4.51$ & $3.91\pm1.00$ & $5.90_{-2.22}^{+2.84}$ & $2.89_{-1.08}^{+1.38}$ & $0.39\pm0.25$ &  \\
\rowcolor{lightgray} &  &  & 21998 & $8.46\pm3.01$ & $6.10\pm2.17$ & $8.82_{-4.40}^{+6.28}$ & $4.29_{-2.14}^{+3.07}$ & $0.39\pm0.34$ &  \\
\rowcolor{lightgray} &  &  & 22911 & $16.65\pm4.26$ & $5.41\pm1.39$ & $7.99_{-3.00}^{+3.81}$ & $3.89_{-1.47}^{+1.88}$ & $0.26\pm0.25$ &  \\
\rowcolor{lightgray} &  &  & 22912 & $10.06\pm3.34$ & $2.83\pm0.94$ & $4.16_{-1.96}^{+2.72}$ & $2.03_{-0.95}^{+1.33}$ & $0.09\pm0.33$ &  \\
NGC-1600 & 04:31:37.34 & -05:04:57.20 & 4283 & $8.72\pm3.01$ & $3.89\pm1.34$ & $3.29_{-1.58}^{+2.28}$ & $1.60_{-0.77}^{+1.11}$ & $0.54\pm0.29$ & GC \\
 &  &  & 4371 & $7.71\pm2.84$ & $2.88\pm1.06$ & $2.45_{-1.25}^{+1.83}$ & $1.19_{-0.61}^{+0.89}$ & $0.52\pm0.32$ &  \\
 &  &  & 22878 & $9.33\pm3.18$ & $2.07\pm0.71$ & $3.02_{-1.45}^{+2.04}$ & $1.47_{-0.71}^{+1.00}$ & $0.41\pm0.31$ &  \\
 &  &  & 21375 & $8.57\pm3.01$ & $2.03\pm0.71$ & $3.17_{-1.55}^{+2.25}$ & $1.54_{-0.75}^{+1.10}$ &  $0.13\pm0.35$ &  \\
\rowcolor{lightgray}NGC-1600 & 04:31:39.82 & -05:04:07.16 & 4283 & $10.00\pm3.15$ & $4.46\pm1.41$ & $3.62_{-1.60}^{+2.29}$ & $1.77_{-0.78}^{+1.11}$ & $0.40\pm0.29$ & DG \\
\rowcolor{lightgray} &  &  & 4371 & $8.00\pm2.83$ & $2.99\pm1.06$ & $5.05_{-2.51}^{+3.67}$ & $2.46_{-1.22}^{+1.79}$ & $0.75\pm0.23$ &  \\
\rowcolor{lightgray} &  &  & 21374 & $6.85\pm2.65$ & $2.66\pm1.03$ & $4.19_{-2.22}^{+3.33}$ & $2.04_{-1.08}^{+1.62}$ & $-0.46\pm0.35$  &  \\
\rowcolor{lightgray} &  &  & 21375 & $14.29\pm3.89$ & $3.38\pm0.92$ & $5.18_{-2.03}^{+2.71}$ & $2.52_{-0.99}^{+1.32}$ & $0.22\pm0.27$ &  \\
\rowcolor{lightgray} &  &  & 21998 & $8.86\pm3.00$ & $6.39\pm2.16$ & $8.96_{-4.22}^{+6.04}$ & $4.36_{-2.05}^{+2.96}$ & $-0.13\pm0.34$  &  \\
\rowcolor{lightgray} &  &  & 22878 & $13.44\pm3.75$ & $2.99\pm0.83$ & $4.55_{-1.83}^{+2.43}$ & $2.22_{-0.89}^{+1.19}$ & $0.15\pm0.28$  &  \\
\rowcolor{lightgray} &  &  & 22911 & $12.86\pm3.61$ & $4.18\pm1.17$ & $6.08_{-2.44}^{+3.29}$ & $2.96_{-1.19}^{+1.60}$ & $-0.22\pm0.27$ &  \\
\rowcolor{lightgray} &  &  & 22912 & $12.70\pm3.61$ & $3.57\pm1.02$ & $5.52_{-2.25}^{+3.03}$ & $2.69_{-1.09}^{+1.48}$ & $-0.21\pm0.28$ &  \\
NGC-1600 & 04:31:40.14 & -05:05:40.90 & 4283 & $12.03\pm4.07$ & $5.36\pm1.81$ & $4.40_{-2.17}^{+2.81}$ & $2.14_{-1.06}^{+1.37}$ & $-0.10\pm0.34$ & DG \\
 &  &  & 4371 & $8.97\pm3.36$ & $3.35\pm1.26$ & $2.78_{-1.49}^{+2.04}$ & $1.35_{-0.72}^{+0.99}$ & $0.61\pm0.29$ &  \\
 &  &  & 21375 & $10.56\pm3.33$ & $2.50\pm0.79$ & $3.84_{-1.71}^{+2.39}$ & $1.87_{-0.83}^{+1.16}$ & $0.30\pm0.30$ &  \\
 &  &  & 22878 & $10.81\pm3.65$ & $2.40\pm0.81$ & $3.54_{-1.72}^{+2.32}$ & $1.72_{-0.84}^{+1.13}$ & $-0.12\pm0.34$ &  \\
 &  &  & 22911 & $11.15\pm3.48$ & $3.62\pm1.13$ & $5.50_{-2.46}^{+3.35}$ & $2.68_{-1.19}^{+1.63}$ & $0.21\pm0.31$ &  \\
 &  &  & 22912 & $15.87\pm4.14$ & $4.46\pm1.16$ & $6.90_{-2.63}^{+3.40}$ & $3.36_{-1.28}^{+1.66}$ & $-0.05\pm0.26$ &  \\
\rowcolor{lightgray}NGC-1600 & 04:31:41.21 & -05:05:42.89 & 4371 & $15.44\pm4.01$ & $5.77\pm1.50$ & $5.01_{-1.89}^{+2.48}$ & $2.44_{-0.92}^{+1.21}$ & $0.74\pm0.17$ & DG \\
\rowcolor{lightgray} &  &  & 22878 & $14.07\pm4.30$ & $3.13\pm0.96$ & $4.58_{-2.05}^{+2.65}$ & $2.23_{-1.00}^{+1.29}$ & $0.17\pm0.30$ &  \\
\rowcolor{lightgray} &  &  & 22912 & $10.72\pm3.32$ & $3.01\pm0.93$ & $4.53_{-1.98}^{+2.77}$ & $2.21_{-0.97}^{+1.35}$ & $0.28\pm0.30$ &  \\
NGC-1600 & 04:31:42.53 & -05:04:36.40 & 21375 & $9.53\pm3.17$ & $2.26\pm0.75$ & $3.57_{-1.68}^{+2.36}$ & $1.74_{-0.82}^{+1.15}$ & $0.65\pm0.27$ & GC \\
 &  &  & 22878 & $6.90\pm2.86$ & $1.53\pm0.64$ & $2.21_{-1.27}^{+1.85}$ & $1.08_{-0.62}^{+0.90}$ &  $1.12\pm0.09$ &  \\
\rowcolor{lightgray}NGC-1600 & 04:31:40.59 & -05:05:07.77 & 22912 & $13.81\pm4.69$ & $3.88\pm1.32$ & $5.59_{-2.82}^{+3.54}$ & $2.72_{-1.37}^{+1.72}$ & $0.23\pm0.32$ & DG/DG \\
NGC-1600 & 04:31:41.52 & -05:04:20.95 & 4238 & $5.60\pm2.68$ & $2.50\pm1.20$ & $2.17_{-1.43}^{+2.14}$ & $1.05_{-0.69}^{+1.04}$ & $0.74\pm0.34$ & DG \\\hline
NGC-7052 & 21:18:36.38 & +26:27:04.43 & 19326 & $19.02\pm4.49$ & $4.93\pm1.16$ & $7.87_{-2.73}^{+3.43}$ & $4.52_{-1.57}^{+2.00}$ & $0.62\pm0.19$ & DG/GC \\
\rowcolor{lightgray} NGC-7052& 21:18:33.77 & +26:27:22.99 & 19326 & $6.00\pm2.45$ & $1.56\pm$0.64 & $2.44_{-1.35}^{+2.13}$ & $1.40_{-0.77}^{+1.23}$ & $0.67\pm0.30$ & DG \\\hline
NGC 7619 & 23:20:11.90 & +08:11:25.11 & 2074 & $39.67\pm6.42$ & $14.98\pm2.42$ & $18.50_{-4.60}^{+5.40}$ & $6.45_{-1.58}^{+1.88}$ & $0.41\pm0.15$ & DG \\
 &  &  & 3955 & $49.99\pm7.31$ & $17.85\pm2.61$ & $15.40_{-3.70}^{+3.70}$ & $5.36_{-1.28}^{+1.29}$ & $0.43\pm0.13$ &  \\
\rowcolor{lightgray}NGC 7619 & 23:20:13.65 & +08:12:50.90 & 2074 & $14.61\pm4.55$ & $5.52\pm1.72$ & $6.66_{-3.07}^{+3.84}$ & $2.32_{-1.07}^{+1.36}$ & $0.23\pm0.29$ & GC \\
NGC 7619 & 23:20:10.43 & +08:11:56.12 & 2074 & $11.73\pm4.34$ & $4.43\pm1.64$ & $5.21_{-2.85}^{+3.60}$ & $1.82_{-0.99}^{+1.26}$ & $0.33\pm0.35$ & GC \\
 &  &  & 3955 & $6.14\pm2.87$ & $2.19\pm1.03$ & $1.97_{-1.28}^{+1.88}$ & $0.69_{-0.45}^{+0.66}$ & $-0.37\pm0.46$ &  \\
\rowcolor{lightgray}NGC 7619 & 23:20:14.40 & +08:11:25.75 & 3955 & $10.80\pm3.94$ & $3.86\pm1.41$ & $3.45_{-1.84}^{+2.40}$ & $1.20_{-0.64}^{+0.84}$ & $0.06\pm0.36$ & GC/GC \\ \hline
\end{tabular}
\end{table*}


\end{document}